\documentclass[apjl,twocolumn]{emulateapj_mod}
\usepackage{epsfig,apjfonts,mathptmx}

\def\gtsima{$\; \buildrel > \over \sim \;$}
\def\ltsima{$\; \buildrel < \over \sim \;$}
\def\prosima{$\; \buildrel \propto \over \sim \;$}
\def\gsim{\lower.5ex\hbox{\gtsima}}
\def\lsim{\lower.5ex\hbox{\ltsima}}
\def\simgt{\lower.5ex\hbox{\gtsima}}
\def\simlt{\lower.5ex\hbox{\ltsima}}
\def\simpr{\lower.5ex\hbox{\prosima}}

\def\h1{$h^{-1}$}
\def\eeq{\end{equation}}
\def\beq{\begin{equation}}
\def\24mu{24\,$\mu{\rm m}$}
\def\70mu{70\,$\mu{\rm m}$}
\def\8mu{8\,$\mu{\rm m}$}

\shorttitle{Star formation laws for disks and starbursts at low and high redshift}

\shortauthors{E. Daddi et al.}

\begin{document}

\title{
Different star formation laws for disks versus starbursts at low and high redshifts
}

   \author{
	  E. Daddi
	  \altaffilmark{1},
          D. Elbaz
          \altaffilmark{1},
          F. Walter
          \altaffilmark{2},
          F. Bournaud
          \altaffilmark{1},
	  F. Salmi
          \altaffilmark{1},
          C. Carilli
          \altaffilmark{3},
          H. Dannerbauer
          \altaffilmark{1},
	  M. Dickinson
	  \altaffilmark{4},
          P. Monaco
          \altaffilmark{5},
          D. Riechers
          \altaffilmark{6}
           }

\altaffiltext{1}{Laboratoire AIM, CEA/DSM - CNRS - Universit\'e Paris Diderot,
       Irfu/Service d'Astrophysique, CEA Saclay, Orme des Merisiers,  91191 Gif-sur-Yvette Cedex, France
    [e-mail: {\em edaddi@cea.fr}]}
\altaffiltext{2}{Max-Planck-Institut f\"ur Astronomie, K\"onigstuhl 17, D-69117 Heidelberg, Germany}
\altaffiltext{3}{National Radio Astronomy Observatory, P.O. Box 0, Socorro, NM 87801}
\altaffiltext{4}{NOAO,  950 N. Cherry Ave., Tucson, AZ, 85719}
\altaffiltext{5}{INAF -- Osservatorio Astronomico di Trieste, Trieste, Italia}
\altaffiltext{6}{Hubble Fellow; Caltech, Pasadena, CA 91125}

\begin{abstract}

   We present evidence that {\it bona fide} disks
   and starburst systems occupy distinct regions in the gas mass versus star formation (SF) rate plane, both
   for the integrated quantities and for the respective surface densities. 
   This result is based on CO observations of galaxy populations at low and high redshifts, and 
   on the current consensus for the CO luminosity to gas mass conversion factors. 
   The data suggest the existence of two different star
   formation regimes: a long-lasting mode for disks and a more
   rapid mode for starbursts, the latter probably occurring
   during major mergers or in dense nuclear SF regions.  Both modes are observable over a large range of SF rates.
   The detection of CO emission from distant near-IR selected galaxies reveals such bimodal
   behavior for the first time, as they allow us to probe
   gas in disk galaxies with much higher SF rates than 
   are seen locally.  The different regimes can potentially be interpreted as the effect 
   of a top-heavy IMF in starbursts. 
   However, we favor a different physical origin
   related to the fraction of molecular gas in dense clouds.
   The IR luminosity to gas mass ratio (i.e., the SF efficiency)
   appears to be inversely proportional to the dynamical (rotation) timescale. Only when
   accounting for the dynamical timescale, a universal SF law is obtained, suggesting
   a direct link between
   global galaxy properties and the local SF rate.

\end{abstract}


\section{Introduction}

Exploring the relation between the gas content and star formation rate (SFR) of galaxies is crucial
to understanding galaxy formation and evolution. This is required to understand the nature
of SF, the parameters that regulate it, and its
possible dependence on local and global galaxy properties 
(e.g., Silk 1997; Elmegreen 2002; Krumholz \& Thompson 2007; McKee \& Ostriker 2007).
In addition, this information is a critical ingredient of
theoretical models of galaxy formation, either based on semianalytical realizations or on numerical
simulations (e.g., Guiderdoni et al.\ 1997; Somerville et al.\ 2001; Monaco et al.\ 2007; Ocvirk et al.\ 2008; Dekel et al.\ 2009; Gnedin et al.\ 2009;
Croton et al.\ 2006). In fact, the physics associated with the conversion of gas into stars inside galaxies
is overwhelmingly complicated, so that theoretical models
generally resort to scaling laws that are calibrated using observations of nearby galaxies.

\citet{schm59} first suggested the existence of
a power law relation between surface densities of SFR and gas masses.
Kennicutt (1998; K98 hereafter) presented a calibration of the Schmidt law with a slope of 1.4 in log space, which has since been the most widely used
in the community. K98 fit the local populations of
spiral galaxies and IR-luminous galaxies (LIRGs/ULIRGs), 
with the gas mass including both neutral (HI) and molecular (H$_2$)
hydrogen for spirals, and molecular gas only for (U)LIRGs 
(as their HI content is likely negligible). The molecular gas
component is routinely estimated using
its most luminous tracer, carbon monoxide (CO), that is generally optically thick. 
This requires an empirical derivation of the conversion factor
to derive molecular gas masses from CO luminosities ($\alpha_{CO}=M_{\rm gas}/L'_{CO}$).
K98 used the Galactic value for all objects in his sample.

\citet{dow98} showed that $\alpha_{CO}$ is smaller for local (U)LIRGs than for
spirals, by about a factor of 6 -- see a detailed discussion and additional references in the review by
\citet{sol05}. \citet{bou07} applied such
 different conversion factors for spirals and (U)LIRGs,
resulting in a steeper relation with a slope of about 1.7. Observations of high redshift submm selected
galaxies (SMGs) can also be fit by the same relation, following \citet{tac08} who argued that a
ULIRG-like $\alpha_{CO}$  is appropriate for SMGs.  
Integrated quantities,
$L'_{\rm CO}$ and IR luminosities ($L_{\rm IR}$)
also appear to follow
a correlation with a slope of 1.7 \citep[e.g.,][]{sol05,gre05}.
In the following discussion, we use a Chabrier (2003) IMF and a standard WMAP cosmology.

   \begin{figure*}
   \centering
   \includegraphics[width=17cm,angle=0]{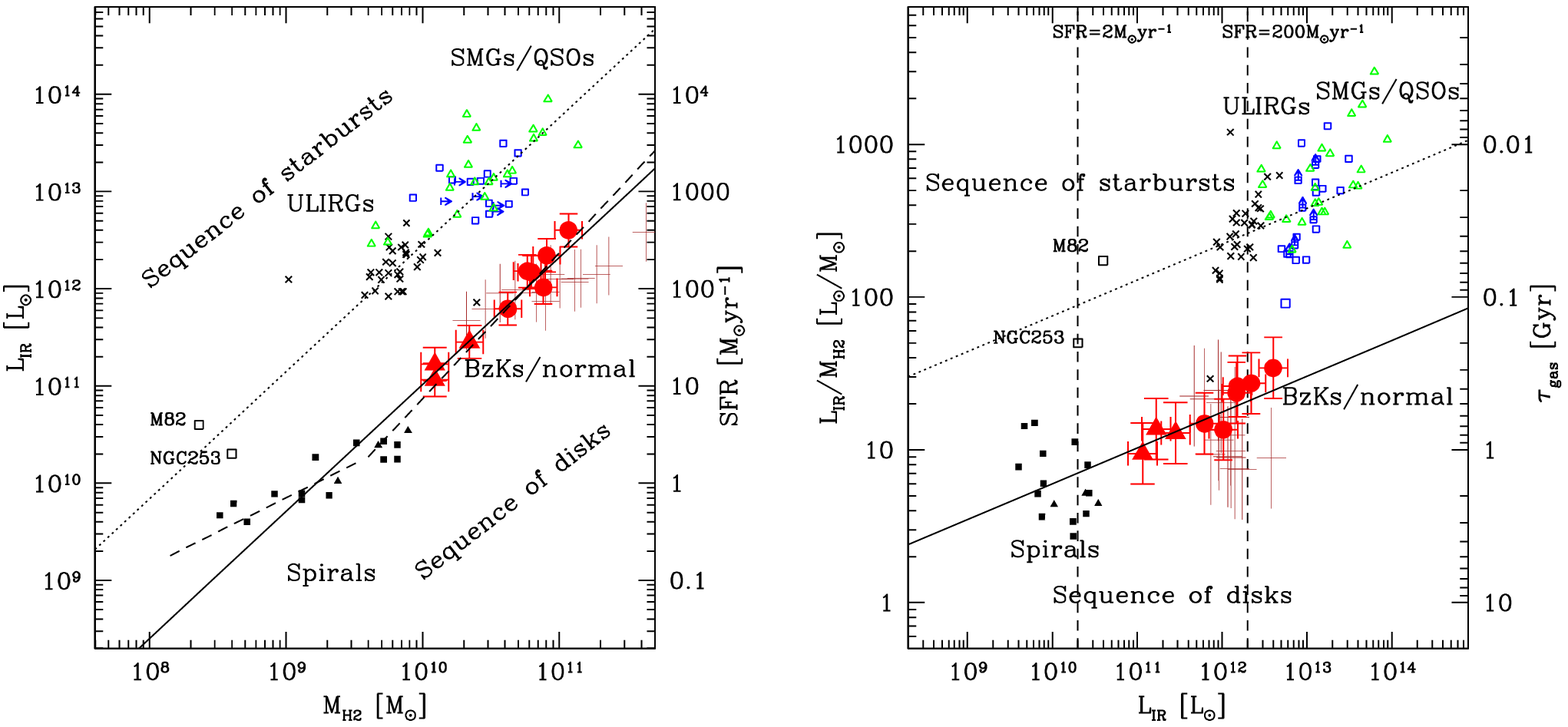}
   \caption{Comparison of molecular gas masses and total
 IR bolometric luminosities:
   BzK galaxies (red filled circles; D10), $z\sim0.5$ disk galaxies (red filled triangles; Salmi et al.\ in preparation),
   $z=1$--2.3 normal galaxies (Tacconi et al.\ 2010; brown crosses),
   SMGs \citep[blue empty squares:][]{gre05,fra08,dad09a,dad09b}, QSOs
   \citep[green triangles: see][]{rie06}, local ULIRGs \citep[black crosses][]{sol97},
      local spirals (black filled squares, Leroy et al.\ 2009; black filled triangles, Wilson et al.\ 2009).
      The two nearby starbursts M82 and the nucleus of NGC~253 are also shown (data from 
      Weiss et al.\ 2001; Houghton et al.\ 1997; Kaneda 2009).
   The solid line (Eq.~1, slope of 1.31 in the left panel) is a fit
   to local spirals and BzK galaxies and the dotted line is the same relation shifted in normalization
   by 1.1~dex. The dashed line in the left panel is a possible double power-law fit to spirals and BzK galaxies.
   For guidance, two vertical lines indicate $SFR=2$ and 200~M$_\odot$~yr$^{-1}$ in the right panel.
           }
              \label{fig:Mgas_scat}%
    \end{figure*}

\section{Relations between IR luminosity and gas masses}

In this letter, we explore the validity of the SF law further by including new observations 
of CO emission in distant near-IR selected galaxies  at $z=0.5$ and $z=1.5$
\citep[][Salmi et al.\ in preparation]{dad08,dad10}.
These allow us to study more typical high-redshift galaxies with star formation rates
much larger than those of local spirals but less extreme than those of distant SMGs.
The sample of 6 CO-detected $z=1.5$ normal (BzK-selected) galaxies is presented in D10. We also use CO detections
of 3 near-IR selected disk galaxies at $z=0.5$. A detailed discussion of the $z=0.5$ dataset will be presented
elsewhere (Salmi et al.\ in preparation). For comparison, we also show
measurements for normal CO-detected galaxies at $z=1$--2.3 from
Tacconi et al.\ (2010), although we do not use these in our analysis.
These new observations are placed in context with literature
data for ULIRGs, SMGs and local samples of disk galaxies.

In order to investigate the location of these populations of normal high-$z$ galaxies 
in the gas mass versus SFR plane, either for the integrated properties or for the surface densities, 
a crucial ingredient is, again, the $\alpha_{\rm CO}$ conversion factor.
Comparing the dynamical and stellar mass estimates,
D10 derive a high 
$\alpha_{CO}=3.6\pm0.8$~$M_\odot$~(K~km~s$^{-1}$~pc$^2$)$^{-1}$\footnote{This conversion factor
refers to the total gas mass, including HI, H$_2$ and Helium, in their proportion within the half light radius.}
for the BzK galaxies,
quite similar to that for local spirals ($\alpha_{CO}=4.6$).
This is not unexpected, given the evidence that the $z\sim1.5$ near-IR selected 
galaxies appear to be high redshift analogs of local disks with enhanced gas content \citep[see e.g., discussions in ][Tacconi et al.\ 2010, and later in this letter]{dad08,dad10,dan09}. 
In the following, we adopt this value of $\alpha_{CO}=3.6$ for the $z=0.5$--2.5 normal galaxies\footnote{Tacconi et al.\ (2010) assume a similar
factor.} and the `consensus'
value for the other populations ($\alpha_{CO}=4.6$ for local spirals, $\alpha_{CO}=0.8$ for local (U)LIRGs and distant SMGs/QSOs),
and  explore the consequences for the relation between gas masses and IR luminosities/SFRs.

   \begin{figure}
   \centering
   \includegraphics[width=8.8cm,angle=0]{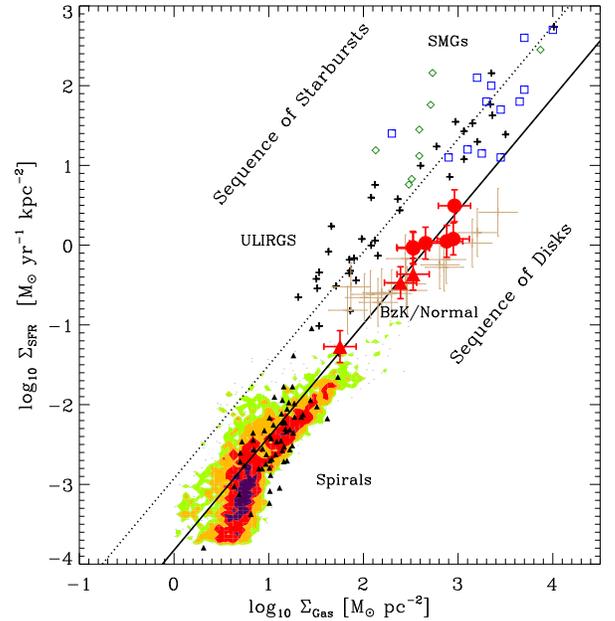}
   \caption{The SFR density as a function of the gas (atomic and molecular) surface density. 
   Red filled circles and triangles are the BzKs (D10; filled) and $z\sim0.5$ disks (Salmi et al.\ in preparation),
   brown crosses are $z=1$--2.3 normal galaxies (Tacconi et al.\ 2010).
   The empty squares are SMGs: \citet{bou07} (blue) and Bothwell et al.\ (2009) (light green). 
   Crosses and filled triangles are (U)LIRGs and spiral
   galaxies from the sample of K98. The shaded regions are THINGS spirals from Bigiel et al.\ (2008).
The lower solid line is a fit to local spirals and $z=1.5$
   BzK galaxies (Eq.~2, slope of 1.42), 
   and the upper dotted line is the same relation shifted up by 0.9~dex  to fit local (U)LIRGs and
   SMGs. SFRs are derived from IR luminosities for the case of a \citet{chab03} IMF.
           }
              \label{fig:KSlaw}%
    \end{figure}

   \begin{figure}
   \centering
   \includegraphics[width=8.8cm,angle=0]{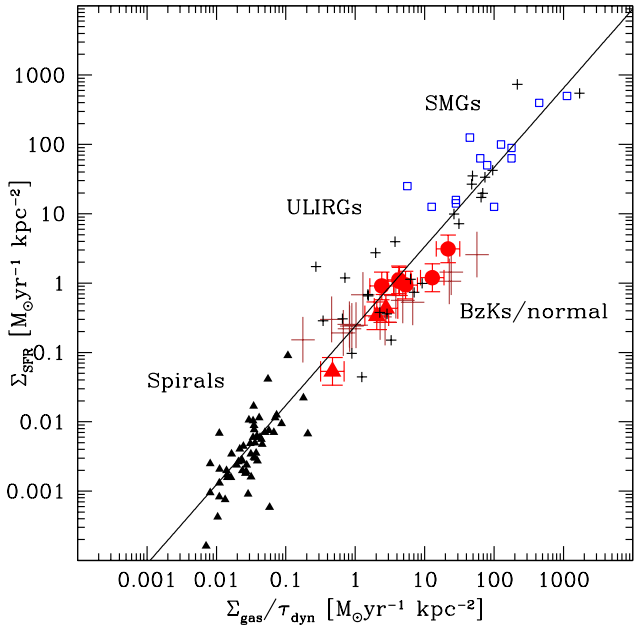}
   \caption{The same as Fig.~\ref{fig:KSlaw}, but with the gas surface densities divided by the dynamical time.
   The best fitting relation is given in Eq.~3 and has a slope of 1.14.
           }
              \label{fig:KSdyn}%
    \end{figure}

Fig.~\ref{fig:Mgas_scat} is equivalent to Fig.~13 in D10, after replacing $L'_{\rm
CO}$ with  $M_{\rm H2}$. 
The right panel shows the ratio of $L_{\rm IR}$ to $M_{\rm H2}$ plotted 
versus  $L_{\rm IR}$. 
The implied gas consumption timescales ($\tau_{\rm gas} = M_{\rm H2}/SFR$; right
panel of Fig.~\ref{fig:Mgas_scat})
are 0.3--0.8~Gyr for the BzK galaxies\footnote{We apply a conversion $SFR[M_\odot$~yr$^{-1}]=10^{-10}\times\,L_{\rm IR}/L_\odot$, treating the two quantities as equivalent. In the
case that a significant AGN contribution affects $L_{\rm IR}$ (e.g., for the QSOs), SFRs would be correspondingly lower.}, 
about 2--3 times that for spirals, and over
one order of magnitude smaller for local (U)LIRGs and distant SMGs. In a simple picture, this finding
can be interpreted in terms of two major SF modes: a long-lasting mode
appropriate for disks, that holds for both local spirals and distant BzK galaxies, and a rapid {\em starburst}
mode appropriate for ULIRGs, local starbursts like M82 or the nucleus of NGC253,
 and distant SMGs/QSOs. 
For the disk galaxies
we formally fit:

\beq
  {\rm log}\ L_{\rm IR}/L_\odot = 1.31\times {\rm log}\ M_{\rm H2}/M_\odot -2.09
\eeq

\noindent
with an error on the slope of 0.09 and a scatter of 0.22~dex. 
Combining ULIRGs and SMGs we find that they define a trend with a similar slope, but with about 10 times
higher $L_{\rm IR}$ at fixed $M_{\rm H2}$.

A similar picture applies to the surface densities (Fig.~\ref{fig:KSlaw}). 
We here use the original K98  measurements for local spirals and (U)LIRGs,
but apply our choice of $\alpha_{\rm CO}$ and a Chabrier (2003) IMF.
For consistency with the K98 relation, we measure $\Sigma_{\rm gas}$ adding
 HI and H$_2$ for spirals, and H$_2$ for IR-luminous galaxies, in Figs~\ref{fig:KSlaw}~and~\ref{fig:KSdyn}. 
The results would not change if we had used H$_2$ only for all galaxies.
Values for SMGs are taken from \citet{bou07}. For the
BzK galaxies we derive gas and SFR surface densities using 
the UV rest frame (SFR) sizes. 
These are consistent with the CO sizes (D10) but are measured at higher S/N ratio. 
Again, we find that the populations 
are split in this diagram and are not well fit by a single sequence. 
Our fit to the local spirals and the BzK galaxies is virtually identical to
the original K98 relation:

\beq
{\rm log} \Sigma_{SFR}/[M_\odot yr^{-1} kpc^{-2}] = 1.42\times {\rm log} \Sigma_{gas}/[M_\odot pc^{-2}]-3.83
\eeq

\noindent
the slope of 1.42 is slightly larger than that of Eq.~1, with an uncertainty of 0.05. 
The scatter along the relation is
0.33~dex. Local (U)LIRG and SMGs/QSOs are consistent with a relation having a similar
slope and normalization higher by 0.9~dex, and a scatter of 0.39~dex

Despite their high SFR$\simgt100$~M$_\odot$~yr$^{-1}$ and $\Sigma_{\rm SFR}\simgt 1
M_\odot$~yr$^{-1}$~kpc$^{-2}$, BzK galaxies are not starbursts, as their SFR can be sustained over timescales
comparable to those of local spiral disks.
On the other hand, M82 and the nucleus of NGC~253 are proto-typical
starbursts, although they only reach a SFR of a few
$M_\odot$~yr$^{-1}$.
Following Figs.~\ref{fig:Mgas_scat}~and~\ref{fig:KSlaw}, and given the $\sim1$~dex displacement of
the disk and starburst sequences, a {\it starburst}
may be quantitatively defined as a galaxy with $L_{\rm IR}$ (or $\Sigma_{\rm SFR}$)
exceeding the value derived from Eqs.~1 (or~2) by more than 0.5~dex.

The situation changes substantially when introducing the dynamical timescale ($\tau_{\rm dyn}$) into the 
picture \citep{sil97,elm02,krum09,ken98}. 
In Fig.~\ref{fig:KSdyn}, we compare $\Sigma_{\rm gas}/\tau_{\rm dyn}$ to $\Sigma_{\rm SFR}$. 
Measurements for spirals and (U)LIRGs are from K98, where $\tau_{\rm dyn}$ is defined to be 
the rotation timescale at the galaxies' outer radius
(although Krumholz et al.\ 2009 use the free fall time).
For the near-IR/optically selected $z=0.5$--2.3 galaxies we evaluate similar quantities at the half light radius. 
Extrapolating the measurements to the outer radius would not affect our
results substantially.  
Quite strikingly, the location of normal high-$z$ galaxies 
is  hardly distinguishable from that of local (U)LIRGs and SMGs. All
observations are well described by the following relation:\\ 

${\rm log} \Sigma_{SFR}/[M_\odot yr^{-1} kpc^{-2}] = $ 
\beq
1.14\times {\rm log} \Sigma_{gas}/\tau_{dyn}/[M_\odot yr^{-1} kpc^{-2}]-0.62.
\eeq 

\noindent
with a slope error of 0.03 and a scatter of 0.44~dex.
The remarkable difference with respect to Figs.~\ref{fig:Mgas_scat}--\ref{fig:KSlaw}
is due to the fact that the normal high-$z$ disk galaxies
have much longer dynamical timescales (given their large sizes) than local (U)LIRGs.

\begin{figure*}
   \centering
   \includegraphics[width=17cm,angle=0]{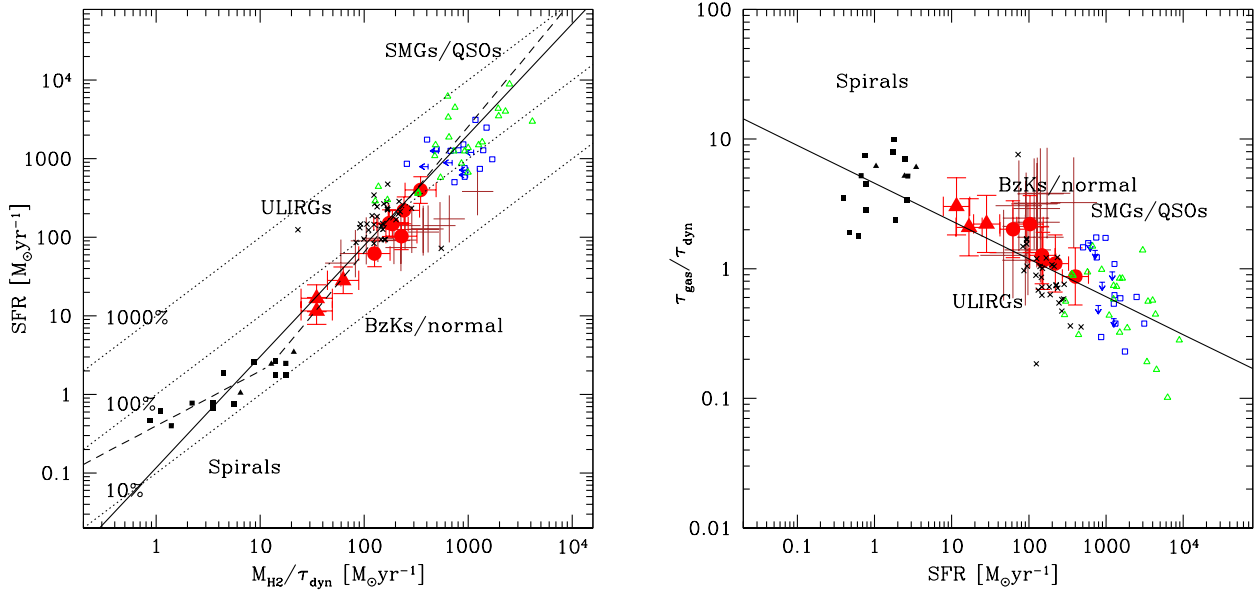}
   \caption{The same as Fig.~\ref{fig:Mgas_scat}, but dividing the gas masses by an average estimate
of the outer radius rotation timescale for the different populations. The dotted lines in the left panel
show constant fractions of gas transformed into stars in each (outer radius) 
rotation. The best fit in the left panel (solid line; Eq.~4) has a slope of 1.42.
           }
              \label{fig:Mgasdyn}%
    \end{figure*}

We can test if this holds also for integrated quantities by dividing the gas masses in 
Fig.~\ref{fig:Mgas_scat} by the
average (outer radius) dynamical timescale in each population. 
Spirals and (U)LIRGs (whose
$\tau_{\rm dyn}$ does not depend on luminosity), have average values of
$\tau_{\rm dyn}=370$~Myr and $\tau_{\rm dyn} = 45$~Myr, respectively  (K98). 
To be compared to $\tau_{\rm dyn} = 33$~Myr for SMGs  \citep{tac06,bou07}. For the QSOs, we use the SMG value.
Assuming a flat rotation curve for BzKs, we get an average $\tau_{\rm dyn}=330$~Myr at the outer radius,
about 3 times longer than at the half light radius, given that for an exponential profile 
90\% of the mass is enclosed within $\sim3$ half light radii.
A similar value is found for our $z=0.5$ disk galaxies and the $z=1$--2.3 objects from Tacconi et al.\,(2010).
Despite this simple approach, Fig.~\ref{fig:Mgasdyn} shows a remarkably tight
trend:
\beq
  {\rm log} SFR/[M_\odot yr^{-1}] = 1.42\times {\rm log} (M_{\rm H2}/\tau_{\rm dyn})/[M_\odot yr^{-1}] -0.86
\eeq

\noindent
with an error in slope of 0.05 and a scatter of 0.25~dex.
Fig.~\ref{fig:Mgasdyn} suggests that roughly 10--50\% of the gas is consumed during each outer disk rotation for local spirals, and some
30--100\% for BzK, $z=0.5$ galaxies and local (U)LIRGs. Some SMGs/QSOs might even consume their gas in less than one rotation.

\section{Discussion} 

The bimodal behavior in Fig.\ref{fig:Mgas_scat}--\ref{fig:KSlaw}
obviously depends on our
assumptions for $\alpha_{\rm\,CO}$. 
Even though we have assumed the most accurate values that are available in the literature, one should keep in mind
that this factor is still relatively poorly constrained. A possible variation of $\alpha_{\rm\,CO}$, e.g., as a function of
galaxy properties, may change  the simple picture outlined here. However, based on our current understanding,
this is unlikely to greatly affect the overall conclusions. Working only with observed quantities, 
the BzK galaxies are already offset 
from local ULIRGs in the $L'_{\rm CO}$/$L_{\rm IR}$ diagram by a factor of 3 (D10).

In summary, the observations of near-IR
selected galaxies at $z=1.5$ do not fit a single relation of $M_{\rm gas}$ versus
$L_{\rm IR}$, neither in terms of integrated quantities nor of surface densities
(Figs.~\ref{fig:Mgas_scat}~and~\ref{fig:KSlaw}). 
There appears to be two main sequences,
one for disks and one for starbursts, with the latter having 10 times higher $L_{\rm IR}$ at fixed $M_{\rm gas}$ 
(either integrated, or for surface densities). 
However, all populations
define a single sequence when dividing the gas masses or surface densities by the respective dynamical timescales.

Fig.~\ref{fig:Mgas_scat}--\ref{fig:KSlaw} cannot be interpreted in terms of a single average trend for all populations
with a large intrinsic scatter. 
Significant {\em systematic} shifts are observed between the BzK and $z=0.5$--2.3 normal galaxies and other highly star
forming galaxies, like local (U)LIRGs and SMGs.
It is worth recalling here that major differences exist in the properties of
local disk galaxies versus local (U)LIRGs, and the same is observed 
(at least on average) between distant near-IR selected BzK galaxies and SMGs 
(see D10, in particular Sect.~7.4,  for a detailed discussion).
Overall, this dichotomy in physical properties of spirals/(U)LIRGs and BzKs/SMGs
makes it less surprising that the two classes define separate sequences in the $M_{\rm
gas}$/$L_{\rm IR}$ plane and correspond to two distinct SF modes.

Previously it was argued that above 100~$M_\odot$~pc$^{-2}$ the gas conversion
factor would   drop by a large factor \citep[see, e.g., the discussion and references in ][]{bou07,tac08}.
This does not seem to be the case for BzK galaxies (D10), that exhibit
SF at 10 times higher gas and SFR surface densities than are seen locally.
One could thus wonder how the same SF mode can be maintained at such high gas surface
densities. It appears that the velocity dispersion is high in these systems \citep[e.g., ][]{for09},
probably due to high turbulence. The implication would be that
BzKs have considerably thicker disks than those of local
spirals. A higher gas filling factor in BzK galaxies 
might also apply.

We now discuss how the observed differences in the $M_{\rm gas}$ versus $L_{\rm IR}$
plane could be interpreted. 
We note that the dichotomy could be altered in the
$M_{\rm gas}$ versus SFR plane if stars were formed with different IMFs in disks and starbursts. For example, if the
$L_{\rm IR}$ to SFR conversion factor was lower for (U)LIRGs and SMGs by a factor of 10, then if we were to plot
SFR instead of $L_{\rm IR}$ in Figs.~\ref{fig:Mgas_scat}--\ref{fig:KSlaw} we would basically end up with a single
SF sequence.
It has been claimed that the IMF in bursts could be top-heavy \citep[e.g.,][]{bau05,elbaz95}.
While the general consensus is that the appropriate
IMF for local spirals is bottom-light as in Chabrier (2003) or \citet{kro02}, very little is 
known observationally for the case of local
(U)LIRGs and even less for high redshift galaxies. In principle, the IMF might explain 
the bimodal behavior at least in part. 

On the other hand, 
we have shown that the ratio of $L_{\rm IR}$ to $M_{\rm gas}$ correlates inversely with the dynamical timescales
of the systems.
The systematic differences disappear once the gas masses are
divided by $\tau_{\rm dyn}$, given that (U)LIRGs and SMGs have much shorter $\tau_{\rm dyn}$ than disk galaxies at $0<z<2.3$. 
This is an important result because, for the first time, we show that global galaxy properties (that determine $\tau_{\rm dyn}$)
are related to the regulation of a local (punctual) process like SF.
The dynamical time is expected to scale with the gas volume  density ($\rho$) as 
$\tau_{\rm dyn}\propto \rho^{-0.5}$ (e.g., Silk et al.\  1997).
This suggests that the different $L_{\rm IR}/M_{\rm gas}$ ratios are not driven
by the IMF (a difference in the IMF would break the single relations found using $\tau_{\rm dyn}$) but by the fact that in some galaxies 
the gas can reach high volume densities and the systems can have
short dynamical timescales so that the fuel is consumed more rapidly
and the resulting SFR per unit of gas mass is higher.
This is consistent with the existence of a unique and approximately
linear correlation between $L_{\rm IR}$ and HCN luminosity
for spirals and local (U)LIRGs \citep{gao04,jun09} and QSOs (with the 
exception of the highest redshift objects; Riechers et al.\ 2007; Gao et al.\
2007). The HCN luminosity is a measure of {\em dense} gas, requiring H$_2$ volume  
densities $>10^4$~cm$^{-3}$, while CO can be thermally  excited for densities $\simlt10^3$~cm$^{-3}$. 
Based on the HCN/$L_{\rm IR}$ relation, the bimodal trends of $L_{\rm IR}$ to $M_{\rm gas}$ in disks versus starbursts 
can be interpreted in terms of a similarly bimodal behavior for the dense gas fractions,
with roughly 10 times higher fraction of dense gas in the starbursts compared to disks at fixed $L_{\rm IR}$
(but also with about twice higher dense gas fractions in BzKs versus local spirals). 
All the observations might thus be explained by a genuine increase of SFR efficiency 
in some galaxy classes,
probably due to the concentration of the gas at high volume  densities.

Major mergers or other kinds of instabilities appear to be the most natural explanation 
for the increased efficiencies in the starbursts (although not all 
mergers will necessarily produce this effect, di Matteo et al.\ 2008).
This is not a new scenario. A higher SF efficiency in merger-driven
starbursts than in rotating disks has often been implemented in
semianalytical models of galaxy formation \citep[e.g.,][]{gui98,som01}, although this has been 
usually interpreted as the
natural outcome of a single SF law with an exponent $>1$, as long as mergers
make gas lose angular momentum and concentrate in the galaxy center. However, in such implementations
the occurrence of very high SF rates in gas-rich disks is
neglected; they thus have a bimodality in surface density, not in the SF law.  
Our analysis suggests that the original K98
calibration can account for the properties of disk galaxies at
low and high redshifts but would underestimate the SF efficiency of
starbursts by a factor of 10. A SF law with a higher exponent of 1.7 (Bouche et al.\ 2007)
would in turn overestimate SF efficiency of gas-rich disks by a similar amount.
An implementation of such a double SF law would surely influence
predictions from semianalytical models of galaxy formation.

The difference in $\alpha_{\rm CO}$ for disks and starbursts helps to {\it hide} 
what we are interpreting here
as large differences in the SF efficiency, expressed in terms of $L_{\rm IR}/M_{\rm gas}$,
by reducing the observed differences in $L_{\rm IR}/L'_{\rm CO}$. There seems to be a conspiracy at work
such that the
particular physical conditions that lead to high $L_{\rm IR}/M_{\rm gas}$ in starbursts also determine variations
in $\alpha_{\rm CO}$ that obscure observationally the differences in SF efficiency.
We emphasize thus that the distinction of starburst (or merging vs non merging) systems is important for 
interpreting CO observations, although this may be difficult on a case by case basis.
We caution that blindly applying the same conversion factor to all high-z observations can lead to confusion. 
Also, care must be
taken that SFR/IR-luminosities are accurately derived. The estimates for BzK galaxies here are based on the
cross-comparison of 3 independent SFR indicators that agree within each other very well, and overall on the global
assessment by Daddi et al.\ (2007ab) on SFR measuraments of near-IR selected galaxies at $z\sim2$. On the other hand, 
purely radio selected or mid-IR selected populations are likely to produce a mixed bag of merging systems
and disk galaxies and can be affected by AGNs. 

\acknowledgements
We thank Adam Leroy for discussions and help with Fig.~2 and Padelis Papadopoulos for discussions.
We acknowledge the funding support of  the ERC-StG grant UPGAL-240039,
ANR-07-BLAN-0228 and ANR-08-JCJC-0008.
DR acknowledges NASA Hubble Fellowship grant
HST-HF-51235.01 awarded by the STScI,
operated by AURA for NASA, contract NAS-5-26555.

\end{document}